\documentclass[%
 reprint,
 amsmath,amssymb,
aip,
]{revtex4-1}

\usepackage{graphicx}
\usepackage{dcolumn}
\usepackage{bm}
\usepackage{graphicx}%
\usepackage[utf8]{inputenc}
 \usepackage{url}
\usepackage{siunitx}

\draft 

\begin{document}

\preprint{APS/123-QED}

\title{High-Efficiency Shallow-Etched Grating on GaAs Membranes for Quantum Photonic Applications}

\author{Xiaoyan Zhou}
\email[]{Author to whom correspondence should be addressed: xiaoyan.zhou@nbi.ku.dk}
\affiliation{Center for Hyrid Quantum Networks (Hy-Q), University of Copenhagen, Blegdamsvej 17, 2100-DK Copenhagen, Denmark}
\author{Irina Kulkova}
\affiliation{Sparrow Quantum, Blegdamsvej 17, 2100-DK Copenhagen, Denmark}
\author{Toke Lund-Hansen}
\affiliation{Sparrow Quantum, Blegdamsvej 17, 2100-DK Copenhagen, Denmark}
\author{Sofie Lindskov Hansen}
\affiliation{Center for Hyrid Quantum Networks (Hy-Q), University of Copenhagen, Blegdamsvej 17, 2100-DK Copenhagen, Denmark}
\author{Peter Lodahl}
\affiliation{Center for Hyrid Quantum Networks (Hy-Q), University of Copenhagen, Blegdamsvej 17, 2100-DK Copenhagen, Denmark}
\author{Leonardo Midolo}
\affiliation{Center for Hyrid Quantum Networks (Hy-Q), University of Copenhagen, Blegdamsvej 17, 2100-DK Copenhagen, Denmark}

\date{\today}

\begin{abstract}
We have designed and fabricated a shallow-etched grating on gallium arsenide nanomembranes for efficient chip-to-fiber coupling in quantum photonic integrated circuits. Experimental results show that the grating provides a fiber-coupling efficiency of $>$60 \%, a greatly suppressed back reflection of $<$1 \% for the designed wavelength of 930 nm, and a 3-dB bandwidth of  $>$43 nm.  Highly efficient single-photon collection from embedded indium arsenide quantum dots to an optical fiber was realized with the designed grating, showing an average sixfold increase in photon count compared to commonly used circular gratings, offering an efficient interface for on-chip quantum information processing. 

\end{abstract}

\pacs{Valid PACS appear here}
                          
\maketitle

Quantum photonic integrated circuits are a promising approach to realizing scalable quantum information processing. Among various materials and platforms, the gallium arsenide (GaAs) material system stands out for offering indispensable quantum functionalities such as single-photon generation, manipulation, and detection, which could be entirely integrated in the same chip\cite{dietrich2016gaas}. In particular, suspended GaAs nanomembranes have been successfully used to construct low-loss and high-performance devices \cite{midolo2015soft,arcari2014near}, including highly-efficient on-demand single-photon sources with embedded indium arsenide (InAs) quantum dots (QDs)\cite{englund2007controlling,lund2008experimental,ba2012enhanced,thyrrestrup2018quantum}, and complex quantum photonic devices such as on-chip phase shifters, routers, and switches\cite{shin2008,luxmoore2013optical,bentham2015chip,midolo2017electro}. 

Despite the blossoming of planar GaAs quantum technologies, efficient device characterization and single-photon extraction remain a challenge due to the lack of a reliable and reproducible method to couple light in and out of the chip. Both in-plane and out-of-plane approaches have been developed aiming at solving this issue. In-plane methods, such as end-fire inverted tapers, spot converters \cite{tran2009photonic,arcari2014near}, and evanescent microfiber couplers\cite{davancco2011efficient,lee2015efficient,daveau2017efficient}, provide the highest coupling efficiencies, however, they are fabrication and alignment demanding. 
Vertical free-space approaches using on-chip gratings are instead easier to implement, more convenient for cryostats having only optical access from a top window, and allow for a quicker characterization of a large number of multi-port devices. 

The most widely used out-coupling structure with QDs in suspended GaAs membranes is a circular grating \cite{faraon2008dipole,luxmoore2013optical,arcari2014near,coles2016chirality,bishop2018electro}. It is a second-order grating with a pitch of $\lambda/(2n)$, where $\lambda$ is the free-space wavelength and $n$ is the refractive index of the membrane. The grating induces destructive interference in the planar propagation direction causing light to be scattered upwards to an objective lens. Such a grating can be fabricated and etched across the membrane with waveguides using a single lithographic step \cite{midolo2015soft} but the poor directionality and non-Gaussian shape of the out-scattered beam result in low coupling efficiency into optical fibers. Moreover, the strong back-reflection of a circular grating adds unwanted Fabry-P\'{e}rot (FP) resonances that ultimately degrade the device performance \cite{coles2016chirality} and make the characterization of integrated circuit (e.g. insertion loss measuements) largely inaccurate. 

In silicon-on-insulator (SOI) photonics, surface focused grating couplers have been developed and optimized for moderately high coupling efficiency (above 50$\%$), large bandwidth, low back-reflection, high polarization selectivity, and a sufficient fiber alignment tolerance\cite{roelkens2010bridging}. These results, obtained by optimizing diffractive gratings over more than a decade in the mature SOI platform, suggest that much can be gained by adapting these structures to the GaAs suspended platform.

In this work we demonstrate a focusing grating coupler integrated with a suspended planar GaAs quantum circuit which includes semiconductor QDs as efficient single-photon sources. Experimental results show a significant improvement of the chip-to-fiber coupling efficiency to $>$60\% and greatly suppressed back-reflection to $<$1\% for the central working wavelength of 930 nm, in good agreement with numerical simulations. Single-photon collection efficiency shows an average sixfold increase compared to circular gratings fabricated on the same sample. These results make shallow-etched gratings suitable for characterizing advanced and sensitive quantum nanophotonic devices, in particular for non-linear transmission experiments \cite{thyrrestrup2018quantum} and insertion loss measurements.

\begin{figure}
\includegraphics{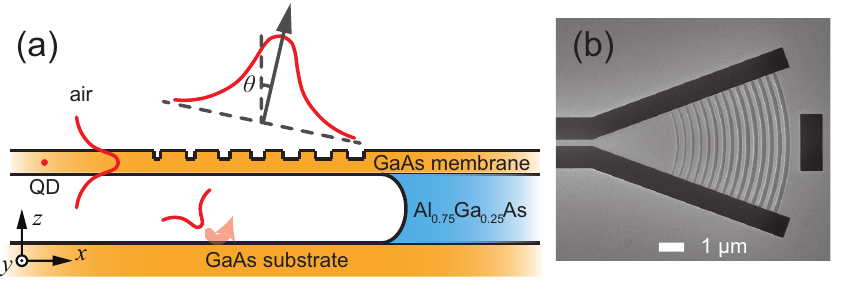}
\caption{\label{fig:concept}(a) Schematic cross-section of the chirped shallow-etched grating couplers integrated with quantum photonic structures in GaAs membrane platform. (b) Top-view scanning electron microscope image of the device.}
\end{figure}

Cross-section schematic of the focusing shallow-etched grating coupler is shown in Fig. \ref{fig:concept}(a). The waveguide mode propagating in the $x$ direction is diffracted upwards by the grating, with an emission angle of $\theta$ to the vertical direction. The basic working principle of the grating can be described by the Bragg condition\cite{chrostowski2015silicon}:

\[ \beta - k_x = m \cdot \frac{2\pi}{\Lambda} \] 
where $\beta = 2\pi \bar{n}_\text{eff} / \lambda_0$ is the propagation constant of the membrane slab mode, $k_x = 2\pi / \lambda_0 \cdot \sin\theta$ is the projected wave-vector component of the diffracted light along the propagation direction, $m$ is the order of diffraction, and $\Lambda$ is the grating period. $\bar{n}_\text{eff}$ is the average effective refractive index of the membrane. Here we consider only the transverse-electric-like (TE-like) mode, because the QDs are in the center of the GaAs membrane and the in-plane dipole moment only couples to the TE-like mode\cite{lodahl2015interfacing}. In the case of the aforementioned circular grating design, $\Lambda$ is chosen so that $\beta = 2\pi / \Lambda$, resulting in the vertical scattering of the first-order diffraction ($k_x$ = 0) and the backward propagation of the second-order diffraction. To avoid back-reflection, a small emission angle ($\theta$) is necessary. 
Using a fixed fill factor FF = 50 \% and an etch depth ED = 50 nm,  our 160 nm GaAs membrane has an estimated $\bar{n}_\text{eff}$ = 2.79. Working wavelength is chosen to be $\lambda_0$ = 930 nm to match the QD emission wavelength. We design the emission angle $\theta = 10^{\circ}$, which gives a pitch $\Lambda = \lambda_0 / (\bar{n}_\text{eff} - \sin \theta)= 355$ nm.

To achieve high coupling efficiency, it is important to ensure constructive interference of the light reflected from the substrate and scattered upwards from the grating. The out-going beam shape should also be optimized to match the fiber mode profile. For this purpose, we used a commercially available finite difference time domain (FDTD) simulation software\cite{fdtd}. First, a uniform grating was optimized for maximum directionality and minimum back-reflection at a wavelength of 930 nm with a two-dimensional model. Using the theoretical parameters calculated above, an optimal sacrificial layer thickness of 1.15 $\mu$m was found to give a maximum transmission of 68.8\% and a reflection of 1.84\%. Then, a full three-dimensional (3D) model of the apodized focusing grating was run to maximize the coupling into a single-mode fiber. We applied the apodization by inducing a linear reduction of the FF along the propagation direction. In fabricated devices, the ED decreases with the FF due to the loading effect in the reactive ion etching (RIE) process. This effect is verified with experimental data and taken into account in the numerical model. 

\begin{figure}
\includegraphics{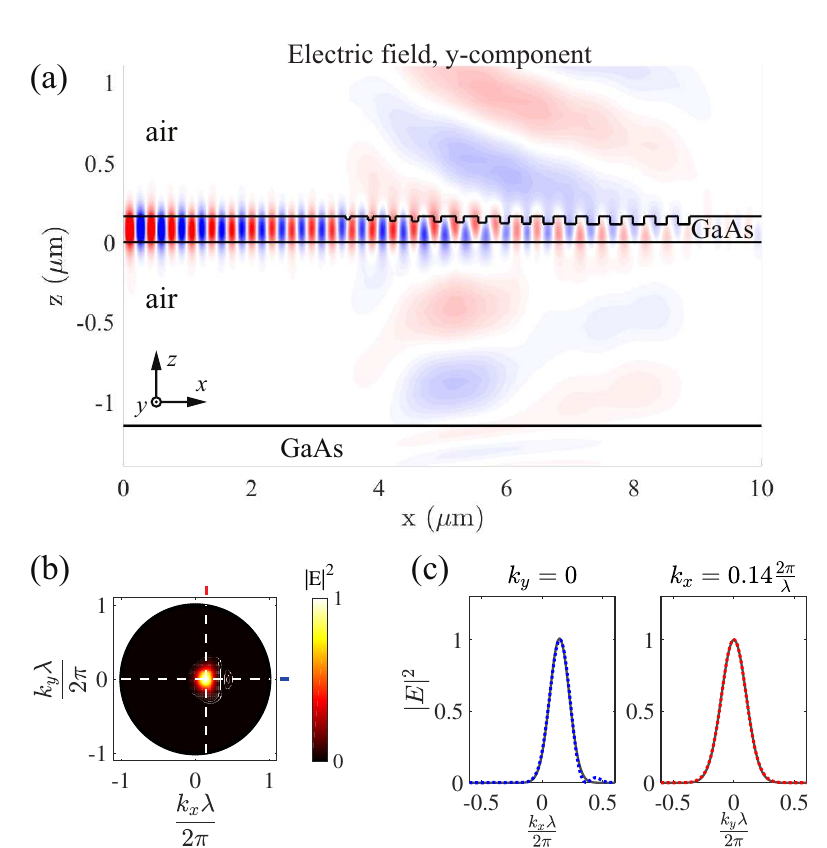}%
\caption{\label{fig:desgin}(a) Cross-section of the 3D optical simulation showing that most of the light is scattered upward with an angle.  (b) The far-field pattern of the grating emission. (c) $x$-cut (blue dotted line) and $y$-cut (red dotted line) of (b) with Gaussian fits shown by solid lines. }%
\end{figure}

Figure \ref{fig:desgin}(a) shows the main electromagnetic field component of the TE-like mode, i.e., $E_y$, in the cross-section of the grating. Most of the light is scattered upward with an emission angle $\theta$ = 8.4$\SI{}{\degree}$ for the designed wavelength of 930 nm. As shown by the far-field plots in Figs. \ref{fig:desgin}(b) and \ref{fig:desgin}(c), the out-going beam has a good directionality and a Gaussian beam profile with small numerical aperture ($\text{NA}_x$ = 0.16, $\text{NA}_y$ = 0.21), facilitating free-space collection and coupling into fibers. 

\begin{figure}
\includegraphics[width=8.5cm]{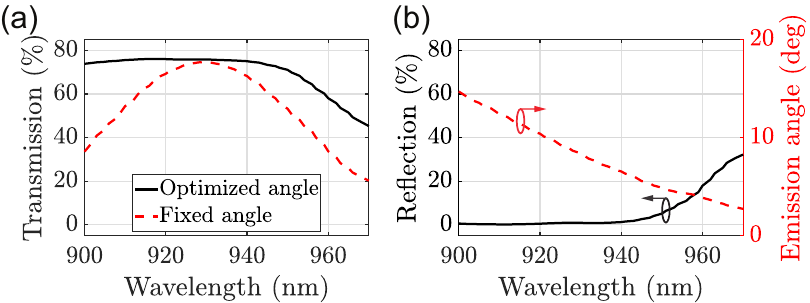}%
\caption{\label{fig:sim}(a) Simulated transmission spectrum to the Gaussian mode at 10 K, considering an adjustable collection angle optimized for each wavelength (continuous line) or a fixed collection angle optimized for 930 nm (dashed line). (b) Simulated reflection increases at shorter wavelengths as the light emission angle goes to zero.}%
\end{figure}

The wavelength-dependent transmission, reflection, and emission angle are shown in Fig. \ref{fig:sim}. At the design wavelength of 930 nm, transmission to fiber reaches 74.6\% with a greatly suppressed reflection of 0.85\%. With fixed collection angle, the grating has a 3-dB transmission bandwidth of 57 nm. We note that the drop in transmission away from the designed wavelength is mainly due to the change of emission angle from the grating. Using an objective with sufficient numerical aperture or controlling the angle between the fiber and the sample, the collection could be optimized at various wavelengths as shown by the solid black line in Fig. \ref{fig:sim}(a). At longer wavelengths, the back-reflection (Fig. \ref{fig:sim}b) increases as the emission angle approaches zero (perfect vertical emission), which can be explained by the increased in-plane component of the second-order diffraction back into the membrane.

The devices were fabricated on a 160-nm-thick GaAs membrane containing a layer of InAs QDs in the center, grown on top of a 1150-nm-thick Al$_{0.75}$Ga$_{0.25}$As sacrificial layer. The layers are grown by molecular beam epitaxy on a (100) GaAs wafer. The fabrication process consists of two electron beam lithography steps followed by dry etching. In the first step, the grating grooves are defined and etched using low-power BCl$_{3}$/Ar RIE. The shallow-etched grooves show a loading effect (RIE lag) which resulted in a size-dependent reduction of the etch rate in the grooves.
In the second step, the waveguides and focusing tapers, aligned within 50 nm tolerance to the grooves, are written and deeply-etched using BCl$_3$/Cl$_2$/Ar inductively coupled plasma RIE (ICP-RIE). To achieve the required alignment tolerance, Ti/Au alignment marks are patterned and lifted off beforehand. Finally, the membrane is released by etching the Al$_{0.75}$Ga$_{0.25}$As layer in a 10\% solution of hydrofluoric acid followed by a residue removal procedure discussed in ref. \cite{midolo2015soft}.

\begin{figure}
\includegraphics[width=8.5cm]{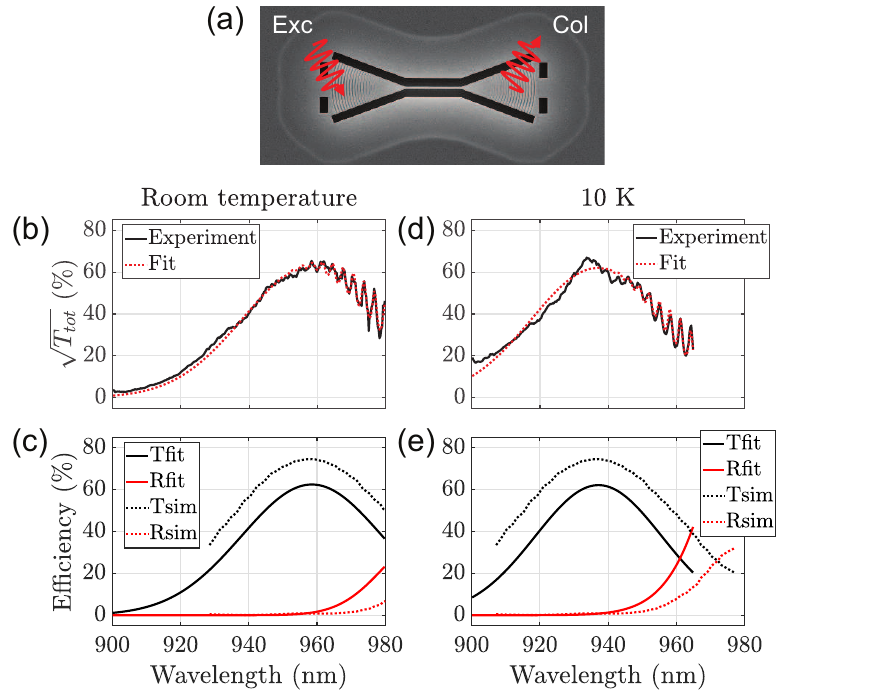}%
\caption{\label{fig:exp}(a) Nanobeam waveguide with in- and out-port gratings for characterization. Experimental and fitted transmission at (b) room temperature and (d) 10 K. The fitted $T$ and $R$ are compared with simulated values at (c) room temperature and (e) 10 K with wavelength-shifted design curves. }%
\end{figure}

To measure the efficiency of the fabricated structures, we use a device consisting of two identical gratings connected by a suspended single-mode nanophotonic waveguide, shown in Fig. \ref{fig:exp}(a). The sample is mounted in a He-flow cryostat and measured at both room and cryogenic (10 K) temperatures. We illuminate one grating from free-space with a broadband laser source (SuperK EXTREME) and collect from the other via a single objective (Nikon 40X, NA=0.6). The collected light is coupled into a fiber through a zoom-fiber collimator (Thorlab ZC618APC-B) with adjustable focal length to match the emission beam size, and sent to a spectrometer. The in- and out-coupling angles from the gratings and the focal length of the fiber collimator are manually adjusted to maximize the transmission signal $I_{tot}$. 
As a reference power for normalization, $I_{ref}$, we use the reflected light from the surface in a homogeneous, unpatterned part of the same sample. The normalized transmission through the whole device is given by $T_{tot} = (I_{tot}/I_{ref})R_{\text{GaAs}}$, where $R_{\text{GaAs}}$ is the reflection coefficient from GaAs bulk, and $R_{\text{GaAs}} = (n_{\text{GaAs}} - 1)^2/(n_{\text{GaAs}} + 1)^2 \approx 0.31$ for normal incidence. 

Neglecting the waveguide loss and assuming that the two gratings are identical, the efficiency of a single grating is $\sqrt{T_{tot}}$, which gives a lower bound estimation of the grating efficiency. The value of $\sqrt{T_{tot}}$ is plotted in Figs. \ref{fig:exp}(b) and \ref{fig:exp}(d) for room and cryogenic (10 K) temperatures, respectively.
The fringes at the long wavelength side of the transmission spectrum indicate back-reflection from the grating, which interferes with the forward propagating light and forms standing waves in the nanobeam waveguide. In this case, a FP cavity model is needed to extract an accurate transmission and reflection spectrum of a single grating. The cavity model of the intensity transmission through the device is :
\[
T_{tot}(\lambda) = \frac{T^2(\lambda)}{|R(\lambda)e^{i2 \beta(\lambda) L}-1|^2}
\]
where $T(\lambda)$ and $R(\lambda)$ are transmission and reflection of a single grating, $\beta(\lambda) = 2 \pi n_{g} / \lambda$ is the propagation constant of the waveguide mode, and $L$ is the length of the waveguide.

We fit the model to the data assuming $T(\lambda)$ and $R(\lambda)$ are described by Gaussian functions, i.e., $T(\lambda) = T_0e^{-((\lambda-\lambda_T)/\sigma_T)^2}$ and $R(\lambda) = R_0e^{-((\lambda-\lambda_R)/\sigma_R)^2}$, 
The fits are shown as dotted red lines in Figs. \ref{fig:exp}(b) and \ref{fig:exp}(d) while the extracted spectra for $T(\lambda)$ and $R(\lambda)$ are shown in Figs. \ref{fig:exp}(c) and \ref{fig:exp}(e). Transmission spectrum with a peak $T_{0}$ = 62.5$\pm$0.3\% at central wavelength $\lambda_T$ = 958.5$\pm$0.1 nm and 3-dB bandwidth of 48.5$\pm$0.6 nm is obtained for room temperature data. At cryogenic temperatures, the transmission peak shifts to shorter wavelengths ($\lambda_T$ = 937.2$\pm$0.2 nm, $T_{0}$ = 62.2$\pm$0.5\%) due to the reduction of material refractive index. The 3-dB bandwidth slightly reduces to 43.7$\pm$0.9 nm. 
The back-reflection at the central wavelength is $<$1\% and negligible for shorter wavelengths as expected from numerical analysis. 
The theoretical $T(\lambda)$ and $R(\lambda)$ spectra are shifted to the experimental central wavelength for ease of comparison. 
The discrepancy between the numerical analysis and the measured data is likely due to loss in the waveguide, fabrication errors, or alignment issues. 

\begin{figure}
\includegraphics[width=8.5cm]{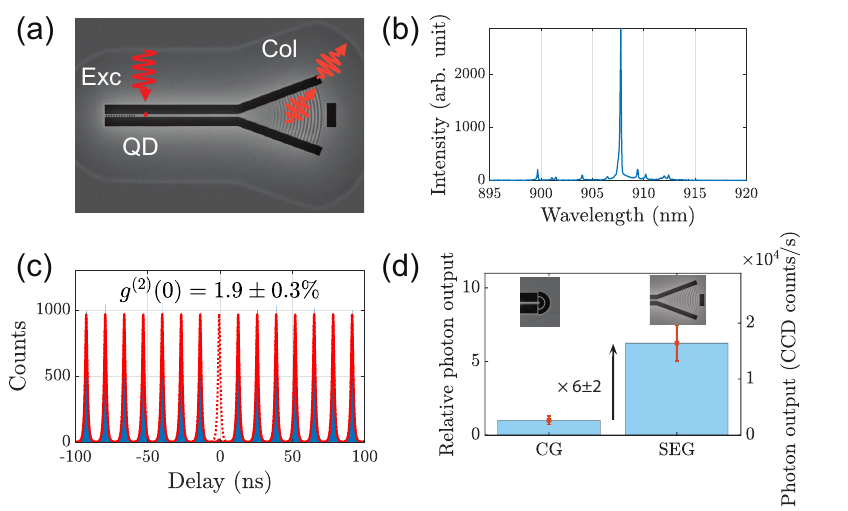}%
\caption{\label{fig:QD}(a) Single-photon source with QDs embedded in nanobeam waveguide terminated by a photonic crystal mirror and a grating. (b) Typical QD emission spectrum. (c) Auto-correlation measurement for the QD emission line at 907.8 nm in (b). The data are fitted (red lines) to extract $g^{(2)}(0)$.  (d) Statistic measurements of collected photon counts from devices with circular grating (CG) and shallow-etched grating (SEG). Error bars represent the standard deviation of the counts. }%
\end{figure}

For the application of efficient single-photon collection, we replace the shallow-etched grating at one end of the waveguide in Fig. \ref{fig:exp}(a) with a photonic crystal (PhC) mirror\cite{sauvan2005modal}, as shown in Fig. \ref{fig:QD}(a). The QDs were excited from free-space by a pulsed laser at 805 nm (above GaAs bandgap) with a repetition rate of 76 MHz. A typical spectrum of QD emission collected from a shallow etched grating device is shown in Fig. \ref{fig:QD}(b). To validate the single-photon nature of the collected light, an auto-correlation measurement shown in Fig. \ref{fig:QD}(c) was performed on a bright emission line at 907.8 nm. The data show a rather low multi-photon emission probability of $g_2(0)$ = 1.9$\pm$0.3\% at zero time delay, indicating a high-level of pure single-photon emission from the device. 

We quantified statistically the collection efficiency improvement to circular gratings by measuring on the nanobeam waveguides with circular gratings. The collected count-rate value from different QDs or devices has a large variation, which can be attributed to various reasons, such as the position-dependence of QD emission efficiency in nanostructures\cite{kirvsanske2017indistinguishable} and QD blinking due to the unstable charge environment in the ungated sample\cite{lodahl2015interfacing}. To compare devices with different output couplers, we measured QD emissions from 10 devices with circular gratings and 15 devices with shallow-etched gratings and select the largest 30\% of the values from each group, i.e., 3 and 5 values from circular gratings and shallow-etched gratings, respectively. The mean value and standard deviation of the count-rates are presented in Fig. \ref{fig:QD}(d), showing a significant increase of the photon emission with a factor of 6$\pm$2, in good agreement with the transmission measurements. The corresponding count-rate of single photons in the optical fiber using shallow-etched gratings is 1.3 MHz (corrected to the spatial filter efficiency of 0.7 and APD efficiency of 0.32). 

To conclude, we have shown a shallow-etched grating coupler on GaAs membrane system for highly efficient out-of-plane fiber coupling. Using a 3D FDTD solver, the gratings are optimized to give 74.6\% transmission to fiber and greatly reduced back-reflection of 0.85\%. We have experimentally measured $>$60\% coupling efficiency with a 3-dB bandwidth of $>$43 nm and $<$1\% reflection on the fabricated structures at both room and cryogenic temperatures, making the designed gratings promising for high performance classical and quantum photonic circuits. As an example, we show that the single-photon collection efficiency from embedded GaAs QDs are greatly improved using the designed grating by a factor of 6 compared to conventional circular gratings on the same chip. The proposed grating enables a fast and accurate characterization of quantum photonic devices and offers an efficient interface for on-chip quantum functional devices.

\begin{acknowledgments}
The authors gratefully acknowledge R{\"u}diger Schott, Andreas D. Wieck, and Arne Ludwig for growing the GaAs wafers with quantum dots, and Tommaso Pregnolato for assistance in fabrication. We gratefully acknowledge financial support from the European Research Council (ERC Advanced Grant 'SCALE'), Innovation Fund Denmark (Quantum Innovation Center 'Qubiz').
\end{acknowledgments}

\bibliography{ShallowEtchedGrating}

\end{document}